\documentstyle[psfig,conf_iap]{article}

\def\eq#1{Eq.~(\ref{eq:#1})}
\def\eqs#1{Eqs.~(\ref{eq:#1})}
\def\eqss#1{(\ref{eq:#1})}

\def\be{\begin{equation}}
\def\ee{\end{equation}}

\def\today{\ifcase\month\or January\or February\or March\or April\or May\or 
June\or July\or August\or September\or October\or November\or December\fi
  \space\number\day, \number\year}

\def\etal{{\it et al.\ }}

\def\eg{{e.g.}}

\def\ltsima{$\; \buildrel < \over \sim \;$}
\def\lsim{\lower.5ex\hbox{\ltsima}}
\def\gtsima{$\; \buildrel > \over \sim \;$}
\def\gsim{\lower.5ex\hbox{\gtsima}}
\def\ga{\mathrel{\hbox{\rlap{\hbox{\lower4pt\hbox{$\sim$}}}\hbox{$>$}}}}
\def\la{\mathrel{\hbox{\rlap{\hbox{\lower4pt\hbox{$\sim$}}}\hbox{$<$}}}}

\def\ifm#1{\relax\ifmmode#1\else$\mathsurround=0pt #1$\fi}
\def\kms{\,{\rm km\,s\ifm{^{-1}}}}
\def\hmpc{\,h\ifm{^{-1}}{\rm Mpc}}

\def\pa{\partial}
 
\def\la{\langle} \def\ra{\rangle}

\def\solar{\ifmmode_{\mathord\odot}\;\else$_{\mathord\odot}\;$\fi}
\def\msun{{\rm M}_{\solar}}

\def\pmb#1{\setbox0=\hbox{#1}%
 \kern-.025em\copy0\kern-\wd0
 \kern.05em\copy0\kern-\wd0
 \kern-.025em\raise.0433em\box0}

\def\vv{\pmb{$v$}}
\def\vx{\pmb{$x$}}

\def\vnabla{\pmb{$\nabla$}}
\def\div{\vnabla\!\cdot\!}

\def\divv{\div\vv}

\def\del{\delta}
\def\delg{g}
\def\delgs{g_{\rm s}}
\def\sig{\sigma}
\def\sb{\sigma_{\rm b}}
\def\sg{\sigma_{\rm g}}

\def\sgs{\sigma_{\rm g,s}}
\def\eps{\epsilon}

\def\cor{{\tt r}}
\def\bh{\hat b}
\def\bt{\tilde b}
\def\bv{b_{\rm var}}
\def\bn{b_{\rm n}}
\def\bp{b_{\rm p}}
\def\av#1{\la #1 \ra}

\def\binv{b_{\rm inv}}
\def\betav{\beta_{\rm var}}
\def\betainv{\beta_{\rm inv}}
\def\betah{\hat\beta}

\def\xigg{\xi_{\rm gg}}
\def\xigm{\xi_{\rm gm}}
\def\ximm{\xi_{\rm mm}}
\def\xibb{\xi_{\rm \eps \eps}}
\def\xibm{\xi_{\rm \eps m}}
\def\xb{r_{\rm b}}
\def\Pgg{P_{\rm gg}}
\def\Pgm{P_{\rm gm}}
\def\Pmm{P_{\rm mm}}

\begin{document}
\medskip

\heading{Galaxy Biasing: Nonlinear, Stochastic and Measurable}\footnote{
to appear in Large-Scale Surveys (IAP Symposium XIV, Paris, France, May 1998),
eds Y. Mellier \& S. Colombi (Editions Frontieres).}

\author{Avishai Dekel}

\noindent{\it Racah Institute of Physics, The  Hebrew University, 
         Jerusalem 91904, Israel}

\begin{abstract}
{\baselineskip 11pt
I describe a general formalism for galaxy biasing \cite{dl98} and its
application to measurements of $\beta$ ($\equiv \Omega^{0.6}/b$),
\eg, via direct comparisons of light and mass and via redshift distortions. 
The linear and deterministic relation $g\!=\!b \del$ between the density 
fluctuation fields of galaxies $g$ and mass $\del$ is replaced by the 
conditional distribution $P(\delg\vert\del)$ of these as random fields, 
smoothed at a given scale and at a given time.  The mean biasing and its
non-linearity are characterized by the conditional mean 
$\av{\delg\vert\del}\!\equiv\!b(\del)\,\del$, and the local scatter 
by the conditional variance $\sb^2(\del)$.
This scatter arises from hidden effects on galaxy formation
and from shot noise.
 
For applications involving second-order local moments, the biasing is 
defined by three natural parameters: the slope $\bh$ of the regression 
of $g$ on $\del$ (replacing $b$), a non-linearity parameter $\bt$, and a 
scatter parameter $\sb$. The ratio 
of variances $\bv^2$ and the correlation coefficient $\cor$ mix these 
parameters.  The non-linearity and scatter lead to underestimates of order 
$\bt^2/\bh^2$ and $\sb^2/\bh^2$ in the different estimators of $\beta$,
which may partly explain the range of estimates.
 
Local stochasticity affects the redshift-distortion analysis only by
limiting the useful range of scales.  In this range, for linear 
stochastic biasing, the analysis reduces to Kaiser's formula for $\bh$ 
(not $\bv$) independent of the scatter.  The distortion analysis is 
affected by non-linearity but in a weak way. 
 
Estimates of the nontrivial features of the biasing scheme are made based 
on simulations \cite{som} and toy models, and a new method for measuring 
them via distribution functions is proposed \cite{sig99}.

}

\end{abstract}

\section{Introduction}
\label{sec:intro}

The fact that galaxies of different types cluster differently 
\cite{dres,lah90,san,lov,her,guzzo} implies that many of them are 
biased tracers of the underlying mass distribution.  
Without such biasing, it is hard to reconcile the existence of
large volumes void of galaxies \cite{kirsh} and the spiky distribution
of galaxies on $\sim 100\hmpc$ scales, today \cite{beks} and 
at high redshifts \cite{ste96,ste98}, with the standard theory of
gravitational instability theory (GI).
There is partial theoretical understanding of the origin of biasing
\cite{kais84,dav,bbks,ds86,dr87,braun,bab,lah92,mo}, supported by
cosmological simulations which confirm the existence of biasing 
\cite{cen,kauf,blan,som} and show that it becomes stronger at high 
redshifts \cite{bag,jing,wech,som}.

The biasing is interesting as a constraint on galaxy formation, but it 
is also of great importance when estimating the cosmological density 
parameter $\Omega$.  If one assumes linear and deterministic biasing  
and applies the linear approximation for GI, $\divv = -f(\Omega)\del$, 
where $f(\Omega)\simeq\Omega^{0.6}$ \cite{peeb80}, the observables 
$g$ and $\divv$ are related via the degenerate combination 
$\beta\equiv f(\Omega)/b$.  Thus, one cannot pretend to have determined 
$\Omega$ by measuring $\beta$ without a detailed knowledge of the 
biasing scheme.

It turns out that different methods lead to different estimates of $\beta$
in the range $0.4 \leq \beta \leq 1.1$ \cite{d94,st,d97,d98}.  
The methods include:
(a) comparisons of local moments of $g$ (from redshift surveys) and $\del$
(from peculiar velocities) or the corresponding power spectra or 
correlation functions; 
(b) linear regressions of the fields $g$ and $\del$, or the corresponding 
velocity fields; and 
(c) analyses of redshift distortions in redshift surveys.
In order to sharpen our determination of $\Omega$, it is important that 
we understand this scatter in $\beta$.  Some of it is due to the different
types of galaxies involved and some may be due to the effects of 
non-linear gravity and perhaps other sources of systematic errors.
Here we investigate the possible contribution of nontrivial properties 
of the biasing scheme such as stochasticity and non-linearity.

The theory of density peaks in a Gaussian random field \cite{kais84,bbks}
predicts that the linear galaxy--galaxy and mass--mass correlation 
functions are related via $\xigg(r) =b^2 \ximm(r)$, where the biasing 
parameter $b$ is a constant independent of scale $r$.  However, a much 
more demanding linear biasing model is often assumed, in which the local 
density fields are related deterministically via the relation 
$\delg(\vx) = b\,\del(\vx)$.  This is not a viable model because
(a) it has no theoretical motivation, (b) if $b>1$ it must break down 
in deep voids because values of $\delg$ below $-1$ are forbidden, 
and (c) conservation of galaxy number implies that the linear biasing 
relation is not preserved during fluctuation growth. Thus, non-linear 
biasing, where $b$ varies with $\del$, is inevitable.
Indeed, the theoretical analysis of the biasing of collapsed halos 
by Mo \& White \cite{mo}, using the extended Press-Schechter approximation 
\cite{bond}, predicts that the biasing is non-linear. It provides a 
useful approximation for its behavior as a function of scale, time and 
mass.  $N$-body simulations, which provide a more accurate description
(see Figure~\ref{fig:1}; \cite{som}), show that this model is indeed
useful.  

Note that once the biasing is non-linear at one smoothing scale, the 
biasing at any other smoothing scale must obey a different functional 
form of $b(\del)$ and is non-deterministic.  Thus, any deviation from 
the simplified linear biasing model must also involve scale-dependence 
and scatter.  Another inevitable source of scatter is physical scatter 
in the efficiency of galaxy formation as a function of $\del$, 
because the mass density at a certain smoothing scale (larger than the 
scale of galaxies) cannot be the sole quantity determining galaxy formation. 
For example, the random variations in the density on smaller scales
and the local geometry of the background structure must play a role too.
These hidden parameters would show up as scatter in the density--density 
relation.  A third obvious source of scatter is the shot noise.  One can 
try to remove it a priori, but this is sometimes difficult because of the 
small-scale anti-correlations introduced by the finite extent of galaxies.
The alternative is to treat the shot noise as an intrinsic part of the local 
stochasticity of the biasing relation.  The scatter arising from all the 
above is clearly seen for halos in simulations including gravity alone 
(\S~\ref{sec:simu}) even before the complex processes of gas dynamics, 
star formation and feedback affect the biasing.

\section{Local Moments: Variances and Linear Regression}
\label{sec:local}



Let $\del(\vx)$ be the field of mass-density 
fluctuations and $\delg(\vx)$ 
the corresponding field of galaxy-density 
fluctuations, at a given time and for a given type of object. 
The fields are both smoothed with a fixed window
which defines the term ``local". 
The local biasing relation is considered to be
a {\it random\,} process, specified by the 
biasing {\it conditional distribution\,} $P(g |\del)$.
Let the one-point probability distribution
functions (PDF) $P(\del)$ and $P(g)$ be of zero means
and standard deviations $\sig^2\equiv\av{\del^2}$ and
$\sg^2\equiv\av{g^2}$.

Define the {\it mean biasing function} $b(\del)$ by the conditional mean,
\be
b(\del)\,\del\equiv\av{g|\del} 
=  \int d g \, P(\delg\vert\del)\, g .
\ee
This function is plotted in Figure~\ref{fig:1}.
It is a natural generalization of the deterministic linear biasing relation, 
$g=b_1\del$.
The function $b(\del)$ allows for any possible {\it non-linear} biasing.
We find it useful to characterize the function $b(\del)$
by its moments $\bh$ and $\bt$ defined by
\be
\bh \equiv\ \av{b(\del)\, \del^2} /\sig^2
\quad {\rm and} \quad
\bt^2 \equiv\ \av{b^2(\del)\, \del^2} /\sig^2 . 
\ee
It will become clear that $\bh$ is the natural extension of $b_1$
and that $\bt/\bh$ is the relevant measure of non-linearity,
independent of stochasticity.

The local {\it statistical\,} character of the biasing can be expressed 
by the conditional moments of higher order about the mean at a given $\delta$. 
Define the random {\it biasing field\,} $\eps$ by
$\eps \equiv g -\av{g |\del}$, with $\la \eps | \del \ra =0$.
The local variance of $\eps$ at a given $\del$ defines the biasing {\it scatter
function\,} $\sb(\del)$ and by averaging over $\del$
one obtains the local biasing {\it scatter parameter}: 
\be
\sb^2(\del) \equiv \av{\eps^2 | \del}/\sig^2, 
\quad 
\sb^2 \equiv \av{\eps^2}/\sig^2 .
\label{eq:sb}
\ee
The scaling by $\sig^2$ is for convenience.
The function $\av{\eps^2 | \del}^{1/2}$ is marked by error bars in
Figure~\ref{fig:1}.
\def\fa{p}
\def\fb{q}
Here and below we make use of a straightforward lemma,
valid for any functions $\fa(g)$ and $\fb(\del)$:
\be
\av{\fa(g)\, \fb(\del)} = \av{\,\av{\fa(g)|\del}_{g|\del}\,\fb(\del)\, }_\del.
\label{eq:lemma}
\ee


From the three basic parameters defined above one can derive other 
biasing parameters. 
A common one is the ratio of {\it variances},
\be
\bv^2 \equiv {\sg^2 / \sig^2} 
= \bt^2 +\sb^2 .
\label{eq:bv}
\ee
The second equality is a result of \eq{lemma}. 
It immediately shows that $\bv$ is sensitive both to non-linearity and
to stochasticity, with $\bv \geq \bt$.
This makes $\bv$ biased compared to $\bh$,  
\be
\bv= \bh\, \left( {\bt^2 / \bh^2} +{\sb^2 / \bh^2} \right)^{1/2} .
\label{eq:bv1}
\ee

Using \eq{lemma}, the mean parameter $\bh$ is related to the
{\it covariance}, 
\be 
\bh\sig^2 = \av{g \del} .
\label{eq:gm}
\ee 
Thus, $\bh$ is the slope of the linear regression of $g$ on $\del$,
which makes it a natural generalization of $b_1$.
Unlike the variance $\sg^2$ in \eq{bv}, the covariance in \eq{gm}
has no contribution from $\sb$.
A complementary parameter to $\bv$ is the linear {\it correlation coefficient},
\be
\cor \equiv {\av{g \del} / (\sg \sig)}
  = {\bh / \bv} 
  =
\left( {\bt^2 / \bh^2} +{\sb^2 / \bh^2} \right)^{-1/2} .
\label{eq:r}
\ee

The ``inverse" regression, of $\del$ on $g$, yields another biasing parameter:
\be
\binv\equiv{\sg^2 / \av{g\del}} 
  ={\bv / \cor}  
  =\bh \left( {\bt^2 / \bh^2} + {\sb^2 / \bh^2} \right) .
\label{eq:binv}
\ee
Thus, $\binv$ is biased relative to $\bh$, even more than $\bv$.
The parameter $\binv$ is close to what is measured in practice
by two-dimensional linear regression \cite{sig98},
because the errors in $\del$ are larger than in $g$.   
Note that $\bt$ and $\sb$ nicely separate the
non-linearity and stochasticity, while $\bv$, $r$
and $\binv$ mix them.

In the case of {\it linear\,} stochastic biasing, 
the above parameters reduce to
\be
\bt=\bh=b_1,\
\bv=b_1 \left( 1+{\sb^2 / b_1^2} \right)^{1/2} ,\
\cor={b_1 / \bv} ,\
\binv=b_1 \left( 1+{\sb^2 / b_1^2} \right) .
\label{eq:lin-sto}
\ee
Thus, $b_1 \leq \bv \leq \binv$.
In the case of non-linear {\it deterministic\,} biasing:
\be
\bt\neq\bh,\quad
\sb=0,\quad
\bv=\bt,\quad
\cor={\bh / \bt},\quad
\binv={\bt^2 / \bh} .
\label{eq:nl-det}
\ee
In the fully degenerate case of linear and deterministic biasing,
all the $b$ parameters are the same, and only then $\cor=1$.

In actual applications,
the above local biasing parameters are involved when the parameter ``$\beta$"
is measured from observational data.
For linear and deterministic biasing this parameter is defined unambiguously
as $\beta_1\equiv f(\Omega)/b_1$, but any deviation from this
model causes us to measure different $\beta$'s 
by the different methods.
For example,
it is $\betav \equiv f(\Omega)/\bv$ which is 
determined from $\sg$ and $\sig f(\Omega)$.
The former is typically determined from a redshift survey,
and the latter either from an analysis of peculiar velocity data,
from the abundance of rich clusters,
or by COBE normalization of a specific power-spectrum shape.
In the case of stochastic biasing $\bv$
is always an overestimate of $\bt$, \eq{bv}, and when the biasing is linear
$\bv$ is an overestimate of $b_1$. Therefore $\betav$ is underestimated 
accordingly.

Another useful way of estimate $\beta$ 
is via the {\it linear regression} of the fields in our cosmological 
neighborhood, \eg, $-\divv(\vx)$ on $\delg(\vx)$
\cite{d93,hud,sig98}.
In the mildly-non-linear regime, $-\divv(\vx)$ is actually replaced by
another function of the first spatial
derivatives of the velocity field, which better approximates the 
scaled mass-density field $f(\Omega)\del(\vx)$ 
\cite{nus}.
The regression is effectively $\del$ on $\delg$, because the errors in
$\divv$ (or $f\del$) are typically more than twice as large as the
errors in $\delg$.
Hence, the measured parameter is close to
$\beta_{\rm inv}\equiv f(\Omega)/\binv$.
In the case of linear and stochastic biasing, \eq{lin-sto}, $\binv$
is an overestimate of $b_1$ so  
the corresponding $\beta$ is underestimated accordingly.

\section{Two-Point Correlations: Redshift Distortions}
\label{sec:cor}

For the analysis of redshift-distortion we need to 
deal with spatial correlations.
Given the random biasing field $\eps$, 
we define the two-point biasing--matter cross-correlation function
and the biasing auto-correlation function by
\be
\xibm(r) \equiv \av{\eps_1 \del_2} ,
\quad
\xibb(r) \equiv \av{\eps_1 \eps_2} , 
\label{eq:xibb}
\ee
where the averaging is over the ensembles at points 1 and 2 separated by $r$.
We define the biasing as {\it local\,} if
$\xibm(r)=0$ for any $r$ and $\xibb(r)=0$ for $r>\xb$, 
where $\xb$ is on the order of the basic smoothing scale.
Using lemmas that are two-point equivalents of \eq{lemma},
one obtains analogous relations to \eqs{gm} and \eqss{bv}.
In the case of {\it linear\,} and {\it local\,} biasing, these become
\be
\xigm(r)=b_1 \ximm(r) ,  \quad
\xigg(r) = b_1^2 \ximm(r) +\xibb(r) ,
\label{eq:xigg_lin}
\ee
Note that the biasing parameter that appears here is $b_1$, not $\bv$.

To see how the power spectra are affected by the biasing scatter,
we approximate the local biasing by a step function:  
$\xibb(r)=\sb^2\sig^2$ for $r<\xb$ and zero otherwise.
Recalling that the power spectra are the Fourier transforms of
the corresponding correlation functions,
we get for $k\ll\xb^{-1}$, from \eq{xigg_lin},
\be
\Pgm(k)= b_1 \Pmm(k) ,
\quad
\Pgg(k)= b_1^2\Pmm(k) + \sb^2\sig^2 V_{\rm b} ,
\label{eq:Padd}
\ee
where $V_{\rm b}$ is the volume associated with $\xb$.
We see that the {\it local\,} biasing scatter adds a constant to $\Pgg(k)$.


We can now proceed to estimating $\beta$ via redshift distortions
\cite{kais87,ham92,ham93,ham97,fish94,heav,cole,fish96,lah96}.
To first order, the local galaxy density fluctuations in redshift space 
($\delgs$) and real space ($\delg$) are related by
$\delgs = \delg - \pa u/\pa r$,
where $u$ is the radial component of the peculiar velocity $\vv$.
Assuming no velocity biasing, linear GI predicts 
$\pa u/\pa r = - \mu^2 f(\Omega) \del$,
where $\mu^2$ is a geometrical
factor depending on the angle between $\vv$ and $\vx$.
Thus, the basic linear relation for redshift distortions is
$g_{\rm s}=g+f\mu^2\del$.  
The general expression for redshift distortions
is obtained from this basic relation by averaging 
$\av{g^s_1 g^s_2}$ over the distributions of $\del$ at a pair of 
points separated by $r$: 
\be
\xigg^s(r) = \xigg(r) + 2(f\mu^2)\, \xigm(r) + (f\mu^2)^2\, \ximm(r) .
\label{eq:dist_xi}
\ee
Recalling that the power spectra are the Fourier transforms of the 
corresponding correlation functions, one can equivalently write an 
expression involving $\Pgg^s(k)$, $\Pgg(k)$, $\Pgm(k)$ and $\Pmm(k)$,
or the analogous spherical harmonics.
 
Next we tie in the biasing scheme.
In the simplified case of linear and deterministic biasing, one simply has
$\Pgg=b_1\Pgm=b_1^2\Pmm$, so the distortion relation reduces to Kaiser's
formula \cite{kais87}, $\Pgg^s=\Pgg(1+\mu^2\beta_1)^2$, 
where $\beta_1\equiv f(\Omega)/b_1$.
In the more realistic case of {\it linear, local}, and {\it stochastic\,} 
biasing, first at zero lag, $\xibb(0)=\sb^2\sig^2$ and 
$\xigg(0)=\bv^2\ximm(0)$. Then, via \eq{bv} and \eq{gm}, the general 
distortion relation, \eq{dist_xi}, reduces to
\be
\sgs^2 = \sg^2 [1 + 2(f\mu^2) \cor \bv^{-1}  + (f\mu^2)^2 \bv^{-2}].
\label{eq:dist_loc_b}
\ee
In this local equation both $\bv$ and $\cor$ are involved in a 
non-trivial way; the distortions depend on the scatter, reflecting
the $\sb^2$ term in \eq{bv}.
On the other hand, at large separations $r>\xb$, where $\xibb$ vanishes,
one obtains instead, from \eq{xigg_lin},
\be
\xigg^s(r) = \xigg(r) [1 + 2(f\mu^2)\, b_1^{-1} + (f\mu^2)^2\, b_1^{-2}] .
\label{eq:dist_xi_lin}
\ee
This is simply the Kaiser formula again, which, unlike \eq{dist_loc_b}, is 
{\it independent\,} of the biasing scatter! 
It involves only the mean biasing parameter $b_1$, in an
expression that is indistinguishable from the deterministic case. 
This is a straightforward result of the assumed locality of the biasing scheme:
the biasing scatter at two distant points is uncorrelated and therefore its
contribution to $\xigg$ cancels out.

The distortion relation for $P(k)$ becomes more complicated because
of the additive term in \eq{Padd}.  For linear biasing, when substituting 
\eq{Padd} in the linear distortion relation, the terms analogous to the ones
involving $b_1^{-1}$ and $b_1^{-2}$ in \eq{dist_xi_lin} for $\xi$
are multiplied by $[1-\sb^2\sig^2 V_{\rm b}/\Pgg(k)]$, a function of $k$.  
The distortion relation for $P(k)$ is thus affected by the biasing 
scatter in a complicated way.
However, there may be a significant $k$ range around the peak
of $P(k)$ in which the additive scatter term is small compared
to the rest. In this range the relation reduces to an expression
similar to \eq{dist_xi_lin} for the corresponding power spectra.
Still, the scatter term always dominates \eq{Padd} at small
and at large $k$'s. 

Equation (7) of Pen \cite{pen}, which involves $\bv$ and $\cor$ like our
\eq{dist_loc_b}, may leave the impression that the redshift-distortion
expression depends on the scatter. In order to obtain his relation
from the general distortion relation, one has to define 
$k$-dependent biasing parameters by $\Pgg(k)=\bv(k)^2\Pmm(k)$
and $\Pgm(k)=\bv(k)\cor(k) \Pmm(k)$.
(Pen's $\beta$ refers to his $b_1$, which is  
our $\bv$, except that he allows it to vary with $k$).
In the case of local biasing, a comparison to our \eq{Padd}
yields $\bv(k)^2 = b_1^2 + \sb^2\sig^2 V_{\rm b}/P_mm(k)$
and $\bv(k) \cor(k) = b_1$.
In the $k$ range near the peak of $\Pmm(k)$, where the constant term
in \eq{Padd} may be negligible, one has $\bv(k) = b_1$ and $\cor(k)=1$,
and there is indeed no sign of the stochasticity in the distortion relation.

While its sensitivity to stochasticity is indirect, 
the redshift distortion analysis {\it is\,} sensitive to
the {\it non-linearity,} of the biasing. A proper analysis
would require a non-linear treatment including
a non-linear generalization of the GI relation $\divv = -f\del$,
because the non-linear effects of biasing and gravity enter at the same
order.
The result is more complicated than \eq{dist_xi}, but is calculable
in principle once one knows the function $b(\del)$ and the one-
and two-point probability distribution functions of $\del$.

\section{Biasing in Simulations and Toy Models}
\label{sec:simu}

In the scheme outlined above, the function $b(\delta)$ contains the 
information about the mean biasing (via the parameter $\bh$) and its 
non-linear features (\eg, via $\bh/\bt$). 
The next quantity of interest in the case
of stochastic biasing is
the conditional standard deviation, the function $\sb(\delta)$,
and its variance over $\del$, $\sb^2$.
In order to evaluate the actual effects of non-linear and stochastic
biasing on the various measurements of $\beta$, one should try
to evaluate these functions or parameters
from simulations, theoretical approximations and observations.

\begin{figure}[t]
{\includegraphics{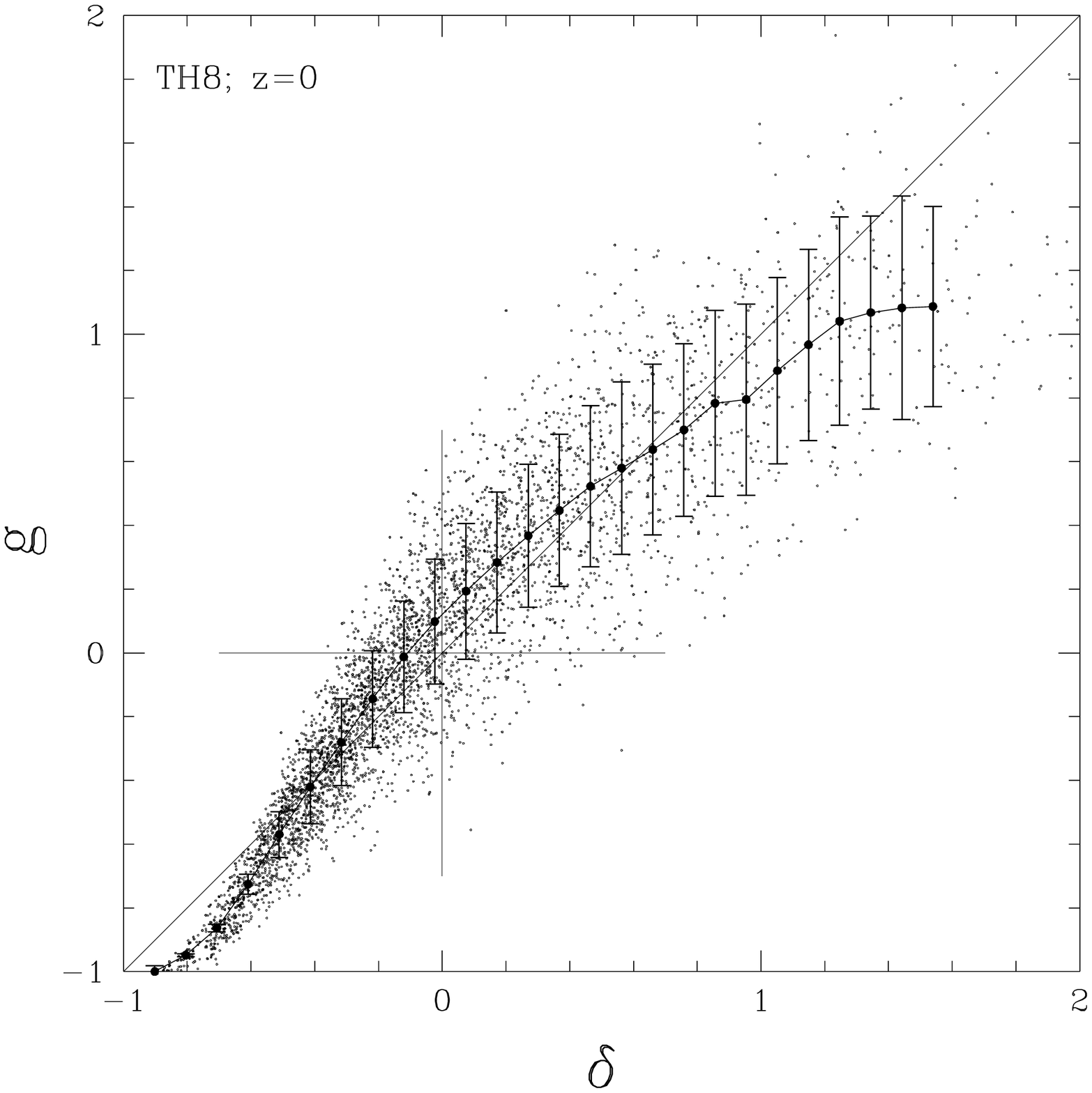}}
{\includegraphics{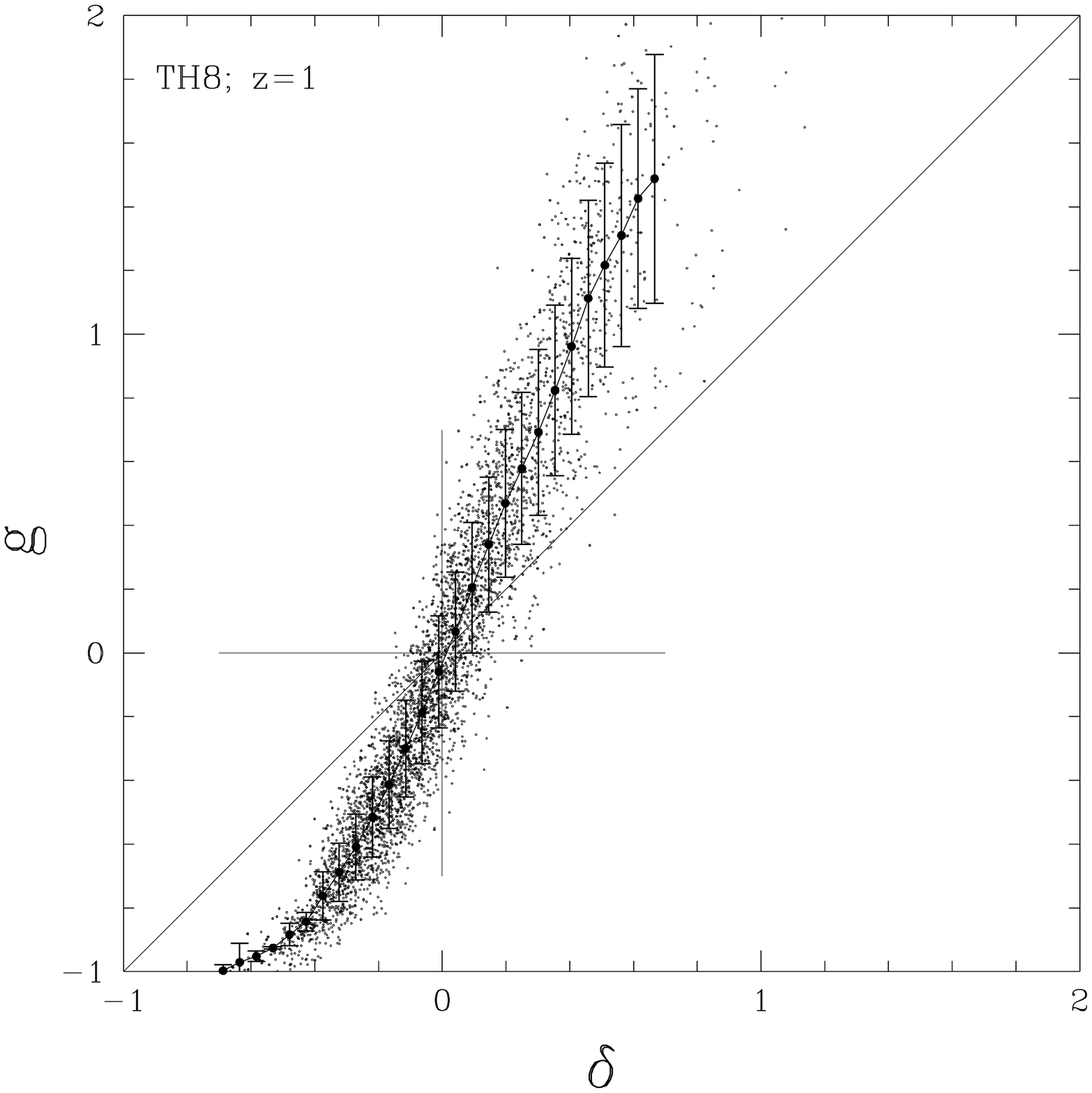}}
\vspace{5.6 cm}
\caption[]{
Biasing of galactic halos versus mass in a cosmological $N$-body simulation,
demonstrating non-linearity and stochasticity.
The conditional mean [$\av{g|\del}=b(\del)\del$] (solid curve) 
and scatter [$\av{\eps^2|\del}=\sb^2(\del) \sig^2$] (error bars) are marked.
The fields smoothed with a top-hat window of radius $8\hmpc$
are plotted at the points of a uniform grid. 
The halos are selected above a mass threshold of $2 \times 10^{12} \msun$. 
Left: at the time when $\sigma_8=0.6$ (\eg, $z=0$).
Right: at an earlier time when $\sigma_8=0.3$ (\eg, $z=1$).
(Based on \cite{som}.)
}
\label{fig:1}
\end{figure}

In an ongoing study that generalizes earlier investigations \cite{cen,mo},
we are investigating the biasing in high-resolution 
$N$-body simulations of several cosmological scenarios, both for galactic
halos and for galaxies as identified using semi-analytic models \cite{som}.
We refer here to a representative cosmological model: 
$\Omega=1$ with a $\tau$CDM power spectrum which roughly obeys
the constraints from large-scale structure.
The simulation mass resolution is $2\times 10^{10} \msun$ inside a box
of comoving side $85\hmpc$. The present epoch is identified with 
$\sigma_8=0.6$.  Figure~\ref{fig:1}  
demonstrates the qualitative features of the biasing scheme.
The non-linear behavior at $\del <1$ is characteristic
of all masses, times, and smoothing scales:
$b(\del)\ll 1$ near $\del=-1$ and it
steepens to $b(\del) > 1$ towards $\del=0$.
At $\del >1$
the behavior strongly depends on the mass, time
and smoothing scale.
The scatter in the figure includes both shot noise and physical scatter
which are hard to separate properly.
In the case shown at $z=0$,
the non-linear parameter is $\bt^2/\bh^2=1.08$,
and the scatter parameter is $\sb^2/\bh^2=0.15$.
The effects of stochasticity and non-linearity in this specific
case thus lead to moderate
differences in the various measures of $\beta$, on the order of
$20-30\%$.
Gas-dynamics and other non-gravitational processes
may extend the range of estimates even further.

Given the distribution $P(\del)$ of the matter fluctuations,
the biasing function $b(\del)$ should obey by definition at least the following 
two constrains. 
First, $g\geq -1$ everywhere, because the galaxy density $\rho_{\rm g}$ 
cannot be negative, with
$g=-1$ at $\del=-1$, because there are no galaxies where there is no matter.
Second, 
$\av{g}=0$ because $g$ describes fluctuations about the mean galaxy density.
An example for a simple functional form that obeys the constraint at
$\delta=-1$ and reduces to the linear relation near $\delta=0$ is
\cite{d93}
\be
\av{g|\del} = c\, (1+\del)^b -1 .
\label{eq:powerb}
\ee
The constraint $\av{g}=0$ is to be enforced by a specific choice of 
the factor $c$ for a given $b$.
With $b>1$, this functional form indeed provides a reasonable fit 
to the simulated halo biasing relation in the $\delta < 0$ regime. 
However, the same value of $b$ does not 
necessarily fit the biasing relation in the $\delta > 0$ regime. 
A better approximation could thus be provided by a combination of two
functions like \eq{powerb}
with two different parameters $\bn$ and $\bp$ in the regimes
$\del\leq 0$ and $\del>0$ respectively. 
The parameter $\bn$ is always larger than unity 
while $\bp$ ranges from slightly below unity to much above unity.
The best fit to Fig.~\ref{fig:1} at $z=0$ has $\bn \sim 2$ and $\bp \sim 1$. 
At high redshift both $\bn$ and $\bp$ become significantly larger.

The non-linear biasing function can alternatively be parameterized by 
\be
\delg = \sum_{n=0}^\infty { b_n \over {n !} } \delta^n .
\label{eq:taylor}
\ee
Since $g$ must average to zero, this general power series can be written as
\be
\av{g|\del} = b_1\del + b_2 (\del^2 -\sig^2)/2
            + b_3 (\del^3 -S)/6 +... ,
\label{eq:taylor3}
\ee
where $\sig^2\equiv\av{\del^2}$, $S\equiv\av{\del^3}$, etc.
This determines the constant term $b_0$.
The constraint at $-1$ provides another relation between the parameters.
Therefore, the expansion to third order contains only two free parameters
out of four.

In order to evaluate the parameters $\bh$ and $\bt$ for these non-linear
toy models, we approximate the distribution $P(\delta)$ as 
log-normal in $\rho/\bar\rho=1+\del$ 
\cite{coles,kof},
where $\sigma$ is the single free parameter.
With \eq{taylor3}, Assuming $b_2\ll b_1$ and $\sig\ll 1$, one obtains 
$
{\bt^2 / \bh^2} \simeq 1 +(1/2) \left({b_2 / b_1}\right)^2 \sig^2
$.
This is always larger than unity, but the deviation is small.
Alternatively, using the functional form of \eq{powerb}, with 
$\bn$ ranging from 1 to 5 and $\bp$ ranging from 0.5 to 3, 
and with $\sig=0.7$, we find 
numerically that $\bt/\bh$ is in the range 1.0 to 1.15.
These two toy models, calibrated by 
the $N$-body simulations, indicate that despite the obvious
non-linearity, especially in the negative regime, the non-linear parameter
$\bt/\bh$ is typically only slightly larger than unity.
This means that the effects of non-linear biasing on measurements of $\beta$
are likely to be relatively small.

\section{Observational Constraints on Biasing}
\label{sec:obs}

Direct constraints on the biasing field should be provided
by the data themselves, of galaxy density (\eg, from redshift surveys)
versus mass density (from peculiar velocity surveys, gravitational
lensing, etc.).
A hint of scatter in the biasing relation
is the fact that the smoothed density peaks of the Great Attractor (GA) 
and Perseus Pisces (PP) are of comparable height in the mass distribution 
as recovered by POTENT from observed velocities 
\cite{d94,dac,d99},
while PP is higher than GA in the galaxy maps
\cite{hud,sig98}.
Another piece of indirect evidence for scatter comes from 
a linear regression of the $1200\kms$-smoothed density fields of POTENT
mass and optical galaxies in our cosmological neighborhood, which
yields a $\chi^2 \sim 2$ per degree of freedom \cite{hud}.
One way to obtain a more reasonable $\chi^2\sim 1$
is to assume a biasing scatter of $\sb\sim 0.5$ 
(while $\sigma \sim 0.3$ at that smoothing).
With $b_1\sim 1$, one has $\sb^2/b_1^2 \sim 0.25$.
This is only a crude estimate; there is yet much to be done with future data
along the lines of reconstructing the ``biasing field" in a given 
region of space.

\def\Cg{C_{\rm g}}
\def\Cd{C_{}}

\begin{figure}[t]
{\includegraphics{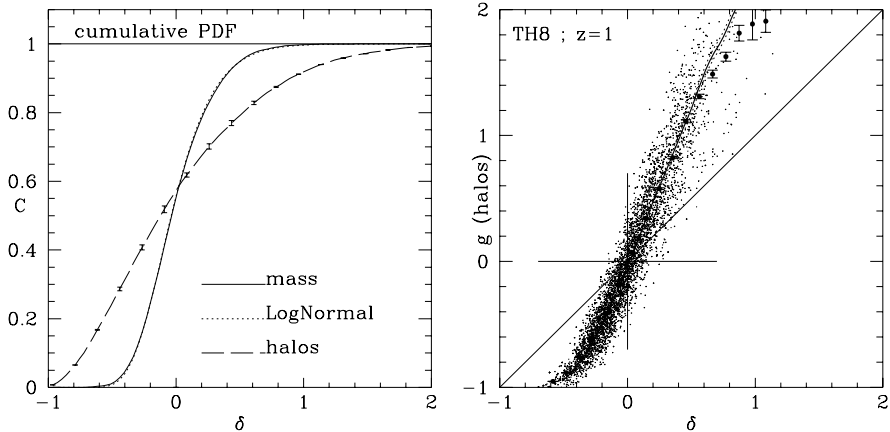}}
\vspace{5.6 cm}
\caption[]{
The PDFs and the mean biasing function, from a cosmological $N$-body
simulation of $\sigma_8=0.3$ ($z=1$)
with top-hat smoothing of $8\hmpc$ and for halos of
$M>2\times 10^{12}\msun$.
Left: the cumulative probability distributions $C$ of density fluctuations 
of halos ($g$) and of mass ($\delta$). 
A log-normal distribution is shown for comparison.
The errors are by bootstrap re-sampling of halos.
The horizontal separation between the curves approximates the mean 
biasing function $\av{g|\del}=b(\del)\sig^2$ 
at the corresponding value of $\delta$. 
Right: the density fields of halos and mass compared at grid points.
The symbols describe the mean biasing function as derived
from this data in bins. 
The solid curve is derived from the PDFs via \eq{pdf}. The dotted
lines mark the error corresponding to the error in the PDF.
(Based on \cite{sig99}.)
}
\label{fig:2}
\end{figure}

We have recently worked out a
promising way to recover the mean biasing function $b(\delta)$ and
its associated parameters $\bh$ and $\bt$ from a measured PDF of
the galaxy distribution \cite{sig99}. 
This method is inspired by a ``de-biasing" technique by
Narayanan \& Weinberg \cite{nar}.
If the biasing relation $g(\delta)$ were deterministic and monotonic,
then it could be derived directly from the cumulative PDFs of
galaxies and mass, $\Cg(g)$ and $\Cd(\del)$, via
\be
g(\del) = \Cg^{-1} [ \Cd (\del)] .
\label{eq:pdf}
\ee
We find, using halos in $N$-body simulations, that this is a good
approximation for $\av{g|\del}$ despite the significant scatter about it.
This is demonstrated in Figure~\ref{fig:2}.

The other key point is that the cumulative PDF of mass density is 
relatively insensitive to the cosmological model or the power spectrum of 
density fluctuations \cite{ber94,ber95}. 
We find \cite{sig99}, using a series of $N$-body simulations
of the CDM family of models in a flat or an open universe
with and without a tilt in the power spectrum,
that, compared to the differences between $\Cg$ and $\Cd$, the latter
can always be properly approximated by a cumulative log-normal distribution 
of $1+\del$ with a single parameter $\sigma$.
Deviations may show up in the extreme tails of the distribution \cite{ber95}, 
which may affect the skewness and higher moments but are of little
concern for our purpose here.
This means that in order to evaluate $b(\del)$
one only needs to measure $\Cg(g)$ from a galaxy density field, 
and add the rms $\sigma$ of mass fluctuations at the same smoothing scale.
Since the redshift surveys are by far richer and more extended than 
peculiar-velocity samples, this method will allow a much better handle
on $b(\del)$ than the local comparison of density fields of galaxies and mass.

\section{Conclusions}
\label{sec:conc}

The key feature in our biasing formalism is the natural separation 
between non-linear and stochastic effects.  The non-linearity is 
expressed by the conditional mean via $b(\del)$, and the  
statistical scatter is measured by the
conditional standard deviation, $\sb(\del)$, and higher moments if necessary.
For analyses using local moments of second order, 
the biasing scheme is characterized by three parameters: 
$\bh$ measuring the mean biasing,
$\bt/\bh$ measuring the effect of non-linearity, 
and $\sb/\bh$ measuring the effect of stochasticity.

Deviations from linear and deterministic biasing typically
result in biased estimates of $\beta$,
which depend on the actual method of measurement.
The non-linearity and the scatter lead to underestimates of order
$\bt^2/\bh^2$ and $\sb^2/\bh^2$ respectively in the different
estimators of $\beta$ relative to $\betah=f(\Omega)/\bh$.
Based on $N$-body simulations and toy models,
the effects of non-linear biasing are typically on the order of 20\%
or less, and the effects of scatter could be larger.
One expects the $\beta$ parameters  
to be biased in the following order: $\betainv < \betav < \betah$.
 
The stochasticity affects the linear redshift-distortion analysis only by
limiting the useful range of scales. 
In this range, the basic expression reduces to the simple
Kaiser formula for $b(\del)=\bh=b_1$ (not $\bv$), and it does not involve the
scatter at all.
The distortion analysis is in principle sensitive to the
non-linear properties of biasing, but they 
are expected to be weak, and of the same order as 
the effects of non-linear GI. 
This is good news for the prospects of measuring an unbiased $\beta$ 
from redshift distortions in the large redshift surveys of the near 
future (2dF and SDSS).

The study of stochastic and non-linear biasing
should be extended to address the {\it time evolution\,} of biasing
because many relevant measurements of galaxy clustering
are now being done at high redshifts.
As seen in Fig.~\ref{fig:1},
the biasing is clearly a strong function of cosmological epoch
\cite{dr87,fry,mo,ste96,ste98,bag,mat,wech,teg,pea98}.
In particular, if galaxy formation is limited to a given epoch
and the biasing is linear, one can show \cite{fry}
that the linear biasing factor $b_1$ would eventually approach unity 
as a simple result of the continuity equation.
Tegmark \& Peebles \cite{teg}
have recently generalized the analytic study of
time evolution to the case of stochastic but still linear biasing 
and showed how $\bv$ and $\cor$ approach unity in this case.
These studies should be extended to the general non-linear case using
our formalism.
Our current simulations  
\cite{som}
are aimed at this goal.
The analysis of simulations could also be extended to include
non-local biasing, using the biasing correlations as defined here.

The PDF (or count in cells) of galaxy density from a large-scale redshift 
survey, plus an estimate of $\sigma$ of
the corresponding mass density, allow a measure of the mean biasing 
function $b(\del)$ and the corresponding non-linearity parameter $\bt/\bh$.
This can be done at low or high redshifts.
Mapping of the biasing field in our cosmological neighborhood,
and estimates of the biasing scatter, are feasible with current 
and future measurements of peculiar velocities 
and careful comparisons to the galaxy distribution. 
The reconstruction of the large-scale mass distribution
based on weak gravitational lensing 
is also becoming promising for this purpose.

In summary,
in order to use the measurements of $\beta$ for an accurate evaluation 
of $\Omega$, one should consider the effects of non-linear and stochastic 
biasing and the associated complications of scale dependence, time 
dependence, and type dependence.
The current different estimates are expected to span a range of 
$\sim 30\%$ in $\beta$ due to stochastic and non-linear biasing.
The analysis of redshift distortions seems to be most promising;
once it is limited to the appropriate range of scales, the analysis is
independent of stochasticity and the non-linear effects are expected to be 
relatively small. The mean biasing function can be extracted from 
the galaxy PDF, and the scatter from theory and local cosmography.

\acknowledgements{
I thank O. Lahav, G. Lemson, A. Nusser, M.J. Rees, Y. Sigad, R.
Somerville, M.A. Strauss, D.H. Weinberg and S.D.M. White 
for collaboration and stimulating discussions.
This research was supported in part by the US-Israel Binational Science
Foundation grant 95-00330, and by the Israel Science Foundation grant 950/95.
}

\def\re{\bibitem}
\def\jeru{in {\it Formation of Structure in the Universe},
     eds.~A. Dekel \& J.P. Ostriker (Cambridge Univ. Press)\ }

\begin{iapbib}{99}{

\re{bab} Babul, A., \& White, S. D. M. 1991, \mn, 253, L31
\re{bag} Bagla, J. S. 1998, \mn, in press (astro-ph/9711081)
\re{bbks} Bardeen, J., Bond, J. R., Kaiser, N., \& Szalay, A.
    1986, \apj, 304, 15
\re{ber94} Bernardeau, F. 1994, \aeta, 291, 697
\re{ber95} Bernardeau, F., \& Kofman, L. 1995, \apj, 443, 479
\re{blan} Blanton, M., Cen, R., Ostriker, J. P., \& Strauss, M. A.
   1998, astro-ph/9807029
\re{bond} Bond, J.R., Cole, S., Efstathiou, G., \& Kaiser, N. 1991, \apj,
   379, 440
\re{braun} Braun, E., Dekel, A., \& Shapiro, P. 1988, \apj, 328, 34
\re{beks} Broadhurst, T. J., Ellis, R. S., Koo, D. C.,
   \& Szalay, A. S., 1990, Nature, 343, 726
\re{cen} Cen, R. Y., \& Ostriker, J. P. 1992, ApJ, 399, L113
\re{cole} Cole, S., Fisher, K. B., \& Weinberg, D. 1995, \mn, 275, 515
\re{coles} Coles, P., \& Jones, B. 1991, \mn, 248, 1
\re{dac} da Costa, L. N., Freudling, W., Wegner, G., Giovanelli, R.,
   Haynes, M. P., \& Salzer, J. J. 1996, \apj , 468, L5
\re{dav} Davis, M., Efstathiou, G., Frenk, C. S., \& White, S. D. M. 1985,
    \apj, 292, 371
\re{d94} Dekel, A. 1994, Ann Rev Ast \& Ast, 32, 371
\re{d98} Dekel, A. 1998, \jeru in press
\re{d93} Dekel, A., Bertschinger, E., Yahil, A., Strauss, M. A., 
  Davis, M., \& Huchra J. P. 1993, \apj , 412, 1 (PI93)
\re{d97} Dekel, A., Burstein, D., \& White, S.D.M.
   1997, in {\it Critical Dialogues in Cosmology}, ed. N. Turok
   (Singapore: World Scientific), p. 175 (astro-ph/9611108)
\re{d99} Dekel, A., Eldar, A., Kolatt, T., Yahil, A., Willick, J. A., Faber,
   S. M., Courteau, S., \& Burstein, D. 1999, in preparation
\re{dl98} Dekel, A., \& Lahav, O. 1998, \apj, submitted (astro-ph/9806193)
\re{dr87} Dekel, A., \& Rees, M. J. 1987, \nat, 326, 455
\re{ds86} Dekel, A., \& Silk, J. 1986, \apj, 303, 39
\re{dres} Dressler, A. 1980, \apj, 236, 351
\re{fish96} Fisher, K. B., \& Nusser, A. 1996, \mn, 279, L1 
\re{fish94} Fisher, K. B., Scharf, C. A., \& Lahav, O. 1994, \mn, 266, 219
\re{fry} Fry, J. N. 1996, \apj, 461, L65
\re{guzzo} Guzzo, L., Strauss, M. A., Fisher, K. B., Giovanelli, R., 
    \& Haynes, M. P.   1997, \apj, 489, 37
\re{ham92} Hamilton, A. J. S. 1992, \apj, 385, L5  
\re{ham93} Hamilton, A. J. S. 1993, \apj, 406, L47  
\re{ham97} Hamilton, A. J. S. 1997, \mn, 289, 285
\re{heav} Heavens, A. F., \& Taylor, A. N. 1995, \mn, 275, 483
\re{her} Hermit, S., Santiago, B. X., Lahav, O., Strauss, M. A., Davis, M., 
     Dressler, A., \& Huchra, J. P. 1996, \mn, 283, 709
\re{hud} Hudson, M. J., Dekel, A., Courteau, S., Faber,
   S. M., \& Willick, J. A. 1995, \mn, 274, 305
\re{jing} Jing, Y. P., \& Suto, Y. 1998, \apj, 494, L5
\re{kais84} Kaiser, N. 1984, \apj, 284, L9
\re{kais87}Kaiser, N. 1987, \mn, 227, 1
\re{kauf} Kauffman, G., Nusser, A., \& Steinmetz, M. 1997, \mn, 286, 795
\re{kirsh} Kirshner, R. P., Oemler, A. Jr., Schechter, P. L., 
   \& Shectman, S. A. 1987 \apj, 314, 493
\re{kof} Kofman, L., Bertschinger, E., Gelb, J., Nusser, A. and Dekel, A.
  1994, \apj , 420, 44
\re{lah96} Lahav, O. 1996, Helvetica Physica Acta, 96, 388  
\re{lah90} Lahav, O., Nemiroff, R. J., \& Piran, T. 1990, \apj, 350, 119
\re{lah92} Lahav, O., \& Saslaw, W. 1992, \apj, 396, 430
\re{lov} Loveday \etal 1995, \apj , 442, 457
\re{nar} Narayanan, V. K., \& Weinberg, D. H. 1998, \apj, in press
      (astro-ph/9806238)
\re{mat} Matarrese, S., Coles, P., Lucchin, F., \& Moscardini, L. 1997, 
      \mn, 286, 115
\re{mo} Mo, H., \& White, S. D. M. 1996, \mn, 282, 347
\re{nus} Nusser, A., Dekel, A., Bertschinger, E., \& Blumenthal,
   G. R. 1991, \apj, 379, 6
\re{pea98} Peacock, J. A. 1998, Phil. Trans. R. Soc. Lond. A, submitted 
   (astro-ph/9805208).
\re{peeb80} Peebles, P. J. E. 1980, {Large-Scale Structure in the Universe}
    (Princeton University Press)
\re{pen} Pen, U.-L. 1998, astro-ph/9711180
\re{san} Santiago, B. X. \& Strauss, M. A. 1992, \apj, 387, 9
\re{sig98} Sigad, Y., Eldar, A., Dekel, A., Strauss, M. A., \& Yahil, A.
   1998, \apj, 495, 516 (astro-ph/9708141)
\re{sig99} Sigad, Y., \& Dekel, A. 1999, in preparation
\re{som} Somerville, R., Sigad, Y., Lemson, G., Dekel, A., 
   Colberg, J., Kauffmann, G., \& White, S. D. M. 1998, in preparation
\re{ste98} Steidel, C. C., Adelberger, K. L., Dickinson, M., Giavalisco, M., 
   Pettini, M., \& Kellogg, M. 1998, \apj, 492, 428 
\re{ste96} Steidel, C. C., Giavalisco, M., Pettini, M., Dickinson, M., \&
   Adelberger, K. L. 1996, \apj, 462, L17
\re{st}Strauss, M. A., \& Willick, J. A. 1995, Phys Rep, 261, 271
\re{teg} Tegmark, M. \& Peebles, P.J.E. 1998, astro-ph/9804067
\re{wech} Wechsler, R. H., Gross, M. A. K., Primack, J. R., Blumenthal, G. R.,
   \& Dekel, A. 1998, \apj, in press (astro-ph/9712141)
}
\end{iapbib}      

\vfill
\end{document}